\documentclass{article}

\usepackage{arxiv}                  

\usepackage[utf8]{inputenc}
\usepackage[T1]{fontenc}
\usepackage{microtype}

\usepackage{amsmath, amssymb, amsthm}
\usepackage{nicefrac}
\usepackage{booktabs}
\usepackage{graphicx}

\usepackage{natbib}
\usepackage{url}
\usepackage{doi}
\usepackage{hyperref}
\usepackage[capitalise]{cleveref}   

\title{Beyond Empirical Bayes: A Hierarchical Bayesian Approach\\
       to Crash Rate Estimation with Missing Traffic Volume}

\author{
    \href{https://orcid.org/0009-0004-1709-8347}{\includegraphics[scale=0.06]{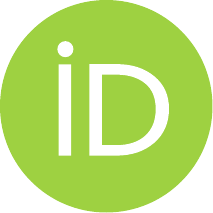}\hspace{1mm}Lars Skaug} \\
    \texttt{larsiskaug@gmail.com}
}

\date{May 2026}

\renewcommand{\shorttitle}{Beyond Empirical Bayes}

\hypersetup{
    pdftitle={Beyond Empirical Bayes: A Hierarchical Bayesian Approach to Crash Rate Estimation with Missing Traffic Volume},
    pdfauthor={Lars Skaug},
    pdfsubject={stat.AP, stat.ME},
    pdfkeywords={Empirical Bayes, Bayesian hierarchical model, crash rate estimation, AADT imputation, road safety, partial pooling},
}

\newcommand{\loocpDelta}{4{,}780}
\newcommand{\loocpDeltaSE}{225}
\newcommand{\loocovDelta}{985}
\newcommand{\loovTwoDelta}{9{,}394}
\newcommand{\loovTwoDeltaSE}{238}
\newcommand{\looPlooVTwo}{1{,}154}
\newcommand{\looKVTwo}{124}
\newcommand{\looPlooSix}{1{,}427}
\newcommand{\looPlooThree}{1{,}466}
\newcommand{\looKSix}{151}

\newcommand{\betaLen}{0.69}
\newcommand{\betaLenLo}{0.68}
\newcommand{\betaLenHi}{0.69}
\newcommand{\betaExpMin}{0.49}
\newcommand{\betaExpMax}{0.70}
\newcommand{\betaExpScalar}{0.58}

\newcommand{\rateInterstate}{1.6}
\newcommand{\rateLocal}{5.9}
\newcommand{\rateOtherArterial}{5.1}

\begin{document}
\maketitle

\begin{abstract}
The Empirical Bayes (EB) procedure of \citet{hauer2002eb} is the workhorse of
highway safety analysis: it combines a Safety Performance Function with
observed crash counts to produce shrinkage estimates of segment-level crash
rates. EB delivers practicality by holding several quantities fixed at
calibration: SPF coefficients, per-type overdispersion, observed ADT, and
a fixed exposure exponent. These assumptions strain when ADT is missing on
a majority of segments. We present a fully Bayesian hierarchical model that
moves beyond EB by relaxing each of these assumptions in a single joint
inference. Fit on Ohio's road inventory (408{,}304 segments, 2.9 million
crashes, 2013--2025), the model jointly imputes missing ADT and estimates
per-segment crash rates with uncertainty. Posterior predictive checks of an
initial fixed-exposure model expose a tail misfit; relaxing the exposure
structure to a per-functional-class exposure exponent and an estimated
length exponent, in place of a single scalar and a fixed offset, resolves
it and improves out-of-sample predictive accuracy (PSIS-LOO
$\Delta\mathrm{elpd} = \loovTwoDelta$, SE $\loovTwoDeltaSE$).
Crash count is sublinear in traffic in every class (exposure exponents
$\betaExpMin$--$\betaExpMax$, all $<1$, the safety-in-numbers effect) and
sublinear in segment length ($\beta_{\mathrm{len}} = \betaLen$). Partial
pooling substantially improves out-of-sample predictive accuracy over
complete pooling (PSIS-LOO $\Delta\mathrm{elpd} = \loocpDelta$, SE
$\loocpDeltaSE$). The Bayesian ADT submodel attains $R^2_{\log} = 0.756$ by encoding county
and functional class as hierarchical priors, versus $0.653$ for a LightGBM
restricted to the same continuous predictors. The output
is a posterior crash rate distribution per segment, replacing the
median-by-type point estimates used in our prior risk-aware routing
framework.
\end{abstract}


\section{Introduction}
\label{sec:intro}

Empirical Bayes (EB), as introduced for highway safety by \citet{hauer2002eb},
combines a Safety Performance Function (SPF) with observed crash counts to
produce shrinkage estimates of segment-level crash rates. Of the many
methods, data sources, and emerging techniques available for crash analysis
\citep{skaug2025review}, EB is the one codified in the AASHTO Highway Safety
Manual and routinely used by state DOTs to identify high-risk segments,
prioritize improvements, and evaluate safety treatments.

EB delivers practicality by holding several quantities fixed at
calibration. The SPF coefficients are point estimates; overdispersion is
one parameter per road type; ADT enters as a known covariate; and the
exposure exponent is fixed during SPF fitting. These choices are deliberate
and appropriate when the assumptions hold. They strain when ADT is missing
on most segments, when road types are heterogeneous within a class, and
when downstream applications such as risk-aware routing require
uncertainty quantification.

This paper presents a fully Bayesian hierarchical model that moves beyond
EB by relaxing those fixed-parameter assumptions in a single joint
inference, enabled by modern MCMC. We fit it on Ohio's road inventory:
408{,}304 segments across 7 functional classes, joined to 2{,}892{,}697
of 2{,}937{,}455 geocoded crashes from 2013--2025 (the rest do not match a
segment). ADT is observed on 128{,}021 segments (31\%) and latent on the
remaining 280{,}283.

The paper builds on \citet{skaug2026cav}, which introduced a risk-aware
routing framework using a frequentist LightGBM ADT model and median-by-type
crash rates. Three limitations identified in that paper motivate the present
work: ADT is imputed without uncertainty propagation; crash rates assume
linear exposure; and rates are computed by simple medians with no borrowing
of strength across similar segments. We address each directly. The present
paper is self-contained: \Cref{sec:eb} gives the EB background needed to
read it without the prior paper, and the routing application is summarized
rather than reproduced.

The contributions are:
\begin{enumerate}
    \item A joint Bayesian hierarchical model for crash counts that
    treats missing ADT as a latent variable, propagates exposure
    uncertainty into rate estimates end-to-end, and estimates the exposure
    exponent from data rather than fixing it at calibration.
    \item An ADT submodel whose hierarchical encoding of county and
    functional class reaches $R^2_{\log} = 0.756$, versus $0.653$ for a
    LightGBM restricted to the same continuous predictors.
    \item A generative-simulation comparison showing that the standard
    two-stage Bayesian multiple-imputation alternative \citep{rubin1987mi}
    introduces material bias in the exposure exponent ($-0.311$) by
    severing the crash-to-ADT likelihood feedback.
    \item Segment-level posterior crash rates with 94\% credible intervals,
    reported here for the measured-ADT segments and obtainable for any of the
    408{,}304 inventory segments through the model's latent-ADT imputation,
    as drop-in replacements for the categorical median rates used in the
    prior routing framework.
\end{enumerate}

\section{Background: Empirical Bayes and Its Fixed Assumptions}
\label{sec:eb}

EB combines two pieces of information about a road segment $i$: a Safety
Performance Function estimate $\mu_i$ for ``similar'' entities (typically
$\mu_i = \alpha \cdot \mathrm{ADT}_i^\beta$ for a given road class), and the
observed crash count $y_i$ on that segment. The EB estimate of the expected
crash count is the weighted average
\begin{equation}
\hat{\lambda}_i^{\mathrm{EB}} = w_i \mu_i + (1 - w_i)\, y_i,
\qquad
w_i = \frac{1}{1 + \mu_i / \phi},
\label{eq:eb}
\end{equation}
where $\phi$ is the negative-binomial overdispersion parameter
\citep{hauer2002eb}. Concretely, a one-mile rural arterial with predicted
$\mu = 4$ crashes/year, observed $y = 12$, and $\phi = 5$ yields $w = 5/9
\approx 0.56$ and $\hat{\lambda}^{\mathrm{EB}} \approx 0.56 \cdot 4 + 0.44
\cdot 12 = 7.5$ crashes/year. The shrinkage pulls rare-event segments toward
their group mean and leaves crash-rich segments closer to their observed
counts.

\Cref{eq:eb} has a Bayesian reading. Under a Gamma$(\mu_i / \phi, 1/\phi)$
prior on the segment's latent rate and a Poisson likelihood for the count,
the posterior mean is exactly $\hat{\lambda}_i^{\mathrm{EB}}$ and $w_i$ is
the posterior weight on the prior. EB and full Bayes share this foundation.
Where they differ is in what is held fixed. EB delivers practicality by
making five such fixed-parameter assumptions:
\begin{enumerate}
    \item The SPF coefficients $(\alpha, \beta)$ are point estimates, fit
    by maximum likelihood at calibration.
    \item The overdispersion parameter $\phi$ is a single number per road
    type.
    \item ADT enters as a known covariate.
    \item The exposure exponent $\beta$ is fixed during SPF calibration,
    often near $0.5$--$0.7$ for rural roads and at $1$ for the simpler
    proportional form.
    \item The output is the point estimate $\hat{\lambda}_i^{\mathrm{EB}}$,
    not a distribution.
\end{enumerate}
Each assumption is appropriate when it holds. Closed-form weights and
type-level SPFs are precisely what makes EB usable in a spreadsheet, and
that accessibility is a feature, not a bug.

The present paper relaxes the five assumptions in a single joint posterior.
The intent is computational evolution rather than correction: EB remains the
right tool where its assumptions hold (we return to this in
\Cref{sec:eb-still-good}). As a preview, the data-estimated exposure
exponents are sublinear in every functional class
($\betaExpMin$--$\betaExpMax$), consistent with Hauer's safety-in-numbers
regime, but pinned to posterior intervals rather than a single chosen value,
and free to differ by class.

\section{Data}
\label{sec:data}

\begin{figure}[t]
    \centering
    \includegraphics[width=\linewidth]{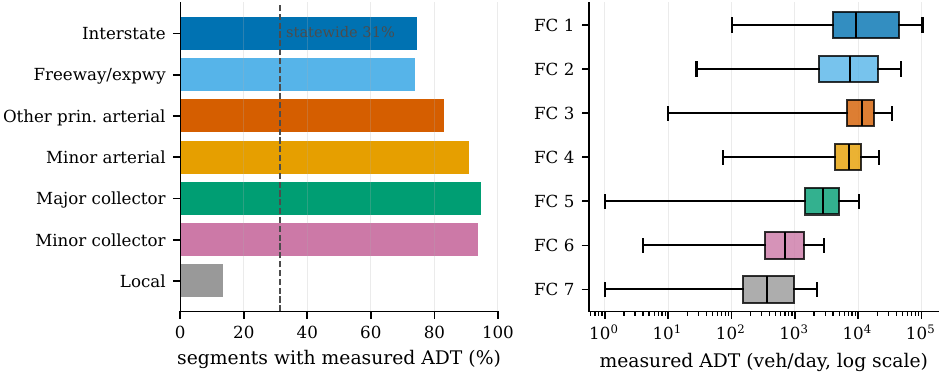}
    \caption{ADT measurement coverage is far from uniform across functional
    class. Left: share of segments with measured ADT by class, against the
    statewide 31\% rate; higher-class roads are nearly fully measured while
    local roads (the bulk of the inventory) are largely not, which is what
    drives the low overall rate. Right: distribution of measured ADT by
    class (log scale), showing the classes separate in level as well as in
    coverage.}
    \label{fig:coverage}
\end{figure}

We use three Ohio DOT datasets: the road inventory (408{,}304 segments
with attributes including functional class, lane count, roadway and
surface widths, posted speed limit, and county); the traffic count segments
(ADT measured on 128{,}021 segments, 31\%, with coverage highly
non-uniform across functional class, \Cref{fig:coverage});
and 2{,}937{,}455 geocoded
crash records from 2013--2025 obtained from ODOT's GCAT export, of which
2{,}892{,}697 ($98.5\%$) match an inventory segment via the linear
reference system (NLF\_ID plus control section) and enter the model; the
unmatched remainder is excluded. The join keeps the geometry and the crash
records in a single coordinate space.

This linear-reference join is the one substantive change from the data
pipeline of \citet{skaug2026cav}, which matched crash points to OSM ways
through a buffer-and-overlap procedure that introduced cross-dataset
alignment errors. Working directly in the inventory's coordinate space
removes that buffer-radius dependence. The broader problem of validating
and correcting such geospatial inventory data is treated in
\citet{skaug2025geospatial}.

The feature set comprises three categorical attributes (functional class,
county, and an urban/rural area code) and four standardized continuous
predictors (number of through lanes, posted speed limit, total roadway
width, and surface width on the left of the centerline). Functional class
has 7 levels; county has 88. The continuous features are drawn from the
inventory's design-attribute fields and are present on every segment.
Functional class anchors both submodels: it groups the SPF and indexes
hierarchical intercepts and slopes. Urban/rural enters the crash submodel
only, to close the lane-count backdoor (urban areas have both more lanes
and higher crash density). Two attributes flagged as predictive in the
prior paper's tree-based EDA are deliberately omitted: route type, which
is collinear with functional class (interstates are FC 1, US highways
FC 2--3, county roads FC 6--7); and route number, whose hundreds of unique
values make it impractical for a Bayesian regression but which is partly
absorbed by the county random effect.

\section{Model}
\label{sec:model}

\subsection{ADT Submodel}
\label{sec:model-adt}

The ADT submodel is a hierarchical log-linear regression. For segment $i$
with measured ADT,
\begin{align}
\log(\mathrm{ADT}_i) &\sim \mathcal{N}(\eta_i^{\mathrm{adt}},\, \sigma_{\mathrm{adt}}), \\
\eta_i^{\mathrm{adt}} &= \mathbf{z}_i^\top \boldsymbol{\beta}_{\mathrm{adt}}
        + \mathbf{z}_i^\top \boldsymbol{\beta}_{\mathrm{adt},\, \mathrm{fc}[i]}^{\mathrm{dev}}
        + \mathbf{z}_i^{(2)\top} \boldsymbol{\gamma}_{\mathrm{adt}}
        + \delta_{\mathrm{county}[i]} + \delta_{\mathrm{fc}[i]},
\end{align}
where $\mathbf{z}_i$ is the vector of standardized continuous predictors
(lanes, speed limit, roadway width, surface width left),
$\boldsymbol{\beta}_{\mathrm{adt}}$ are global slopes,
$\boldsymbol{\beta}_{\mathrm{adt}, \mathrm{fc}}^{\mathrm{dev}}$ are FC-level
slope deviations under a non-centered parameterization,
$\mathbf{z}_i^{(2)}$ holds the squared standardized predictors for
quadratic curvature, and $\delta_{\mathrm{county}}, \delta_{\mathrm{fc}}$
are random intercepts. Priors are weakly informative:
$\boldsymbol{\beta}_{\mathrm{adt}} \sim \mathcal{N}(0, 5)$,
$\sigma_{\mathrm{adt}} \sim \mathrm{HalfNormal}(2)$,
$\sigma_{\mathrm{county}} \sim \mathrm{HalfNormal}(2)$,
$\sigma_{\mathrm{fc}} \sim \mathrm{HalfNormal}(5)$. The wider hyperprior on
$\sigma_{\mathrm{fc}}$ was chosen after an initial run with
$\mathrm{HalfNormal}(1)$ produced a posterior concentrating at $\sim 4.4$
and severe convergence failure; this is recorded honestly because the
correction is itself a useful artefact for replicators.

We compare this submodel to a LightGBM regressor using identical 5-fold
out-of-fold splits. Given the same four continuous features, the Bayesian
model attains $R^2_{\log} = 0.756$ (residual standard deviation $0.850$ on
the log scale) against $0.653$ ($1.014$) for LightGBM. This is not an
equal-information comparison: the Bayesian model additionally encodes county
and functional class through its hierarchical intercepts, which the
continuous-only tree cannot see. When the prior paper gave LightGBM those
high-cardinality categoricals (county, route type) directly, among $8$
features, it reached $R^2_{\log} = 0.877$. The point is therefore not that
the Bayesian model dominates a boosted tree, but that partial pooling is an
efficient way to use the categorical hierarchy: it recovers through a prior
most of the signal a tree extracts through deep splits, and carries that
signal as uncertainty that propagates into the joint model.

\subsection{Joint Crash + ADT Model}
\label{sec:model-joint}

The crash submodel is a negative binomial regression for the count $y_i$:
\begin{align}
y_i &\sim \mathrm{NegBinom}(\mu_i,\, \alpha_{\mathrm{fc}[i]}), \\
\log \mu_i &= \beta_{0,\, \mathrm{fc}[i]}
        + \beta_{1,\, \mathrm{fc}[i]} \cdot \mathrm{lanes}_i^z
        + \beta_2 \cdot \mathrm{urban}_i
        + \beta_3 \cdot \mathrm{speed}_i^z
        + \beta_4 \cdot \mathrm{rdwy}_i^z
        + \beta_5 \cdot \mathrm{surf}_i^z \notag \\
        &\quad + \log(365 \cdot T)
        + \beta_{\mathrm{len}} \cdot \log(L_i)
        + \beta_{\mathrm{exp},\, \mathrm{fc}[i]} \cdot \log(\mathrm{ADT}_i),
\label{eq:crash}
\end{align}
where $L_i$ is segment length in miles, $T = 13$ years, and $z$
superscripts denote standardization. Group intercepts and lane slopes are
partially pooled: $\beta_{0,\, \mathrm{fc}} \sim \mathcal{N}(\mu_{\beta_0},
\sigma_{\beta_0})$ and $\beta_{1,\, \mathrm{fc}} \sim \mathcal{N}(\mu_{\beta_1},
\sigma_{\beta_1})$, with the centered parameterization (appropriate for
$7$ well-populated FC groups). The exposure exponent is likewise partially
pooled across functional class, $\beta_{\mathrm{exp},\, \mathrm{fc}} \sim
\mathcal{N}(\mu_{\beta_{\mathrm{exp}}}, \sigma_{\beta_{\mathrm{exp}}})$ with
$\mu_{\beta_{\mathrm{exp}}} \sim \mathcal{N}(1, 0.3)$, centered at
proportionality but admitting a sublinear posterior, and free to differ by
class. Per-FC overdispersion $\alpha_{\mathrm{fc}} \sim \mathrm{HalfNormal}(2)$.
Fixed-effect slopes have priors $\beta_{2..5} \sim \mathcal{N}(0, 0.5)$ and
the length exponent $\beta_{\mathrm{len}} \sim \mathcal{N}(1, 0.3)$. The
intercept hyperprior $\mu_{\beta_0} \sim \mathcal{N}(-13, 2)$ is a
deliberately weak anchor: combined with the exposure terms, a center of
$-13$ places the implied prior crash counts on the order of a few events per
segment over the 13-year window, the right order of magnitude for the
observed range across functional classes, while the $\pm 2$ prior
standard deviation keeps the model from committing to that location and
lets the likelihood move it.

Both exposure terms in \cref{eq:crash}, length and ADT, carry
estimated exponents rather than fixed ones. The natural prior expectation
differs between them: segment length is a geometric extent, so doubling a
segment's length might be expected to double the spatial opportunity for a
crash ($\beta_{\mathrm{len}} \approx 1$), whereas ADT is a temporal
interaction rate for which \citet{hauer2001overdispersion} documents a
sublinear effect: doubling traffic does not double conflicts because gaps
shrink and drivers adjust. We accordingly center both priors at
proportionality but estimate both exponents; \cref{sec:ppc} shows that the
data revise \emph{both} downward, and that the exposure exponent is not even
constant across functional classes. This relaxation, treating the
exposure structure as estimated and class-varying rather than fixed, is the
same move the model already makes for the intercept and lane slope, and we arrived
at it by model checking rather than by assumption (\cref{sec:ppc}).

ADT enters the crash likelihood as observed data where measured and as a
latent variable where not. For latent ADT, the prior is the ADT submodel
posterior predictive given the segment's covariates; crash counts then
feed back into latent ADT through the joint likelihood. A segment with
surprisingly many crashes is pulled toward a higher latent ADT (more
exposure required to explain the count), which in turn informs the global
$\beta_{\mathrm{exp}}$.

Severing this feedback, as a two-stage approach does, removes the
very signal that pins the exposure exponent. The mechanism is
identification. On a segment with missing ADT, the only data bearing on
its exposure are its covariates and its own crash count. Under joint
inference, $\beta_{\mathrm{exp}}$ is identified by the covariation between
latent ADT and crash counts across such segments: a high count pulls the
latent ADT up, and the magnitude of that pull \emph{is} the exponent. A
two-stage pipeline instead imputes each latent ADT from the ADT submodel
alone (covariates only) and then freezes it, so the crash likelihood can
no longer revise ADT to be consistent with the count it must explain. The
frozen imputations are then noisy proxies for the true latent exposure,
and a count regression on a noisy regressor attenuates its slope toward
zero (classical errors-in-variables), with the intercept rising to keep
the mean prediction in place. We confirm this mechanism on simulated data
below, where the attenuation is strong enough to bias
$\beta_{\mathrm{exp}}$ downward even against a prior centered above the
true value.

\subsection{Why Joint, Not Two-Stage: A Simulation}
\label{sec:simulation}

Before fitting on real data we ran a generative simulation
\citep{gelman2020workflow}. With $N = 5{,}000$ segments, $5$ functional
classes, $10$ counties, and parameter values chosen near the posteriors
we anticipated, we masked $70\%$ of ADT to mimic the real missingness
fraction and fit two competing pipelines on the same simulated data:
(i) Bayesian multiple imputation \citep{rubin1987mi}, fitting the ADT
submodel separately, drawing $S = 5$ posterior predictive imputations,
fitting the crash model per imputation, and pooling traces; and
(ii) the joint model in which both submodels share inference and the
crash likelihood feeds back into the latent ADT.

\begin{figure}[t]
    \centering
    \includegraphics[width=\linewidth]{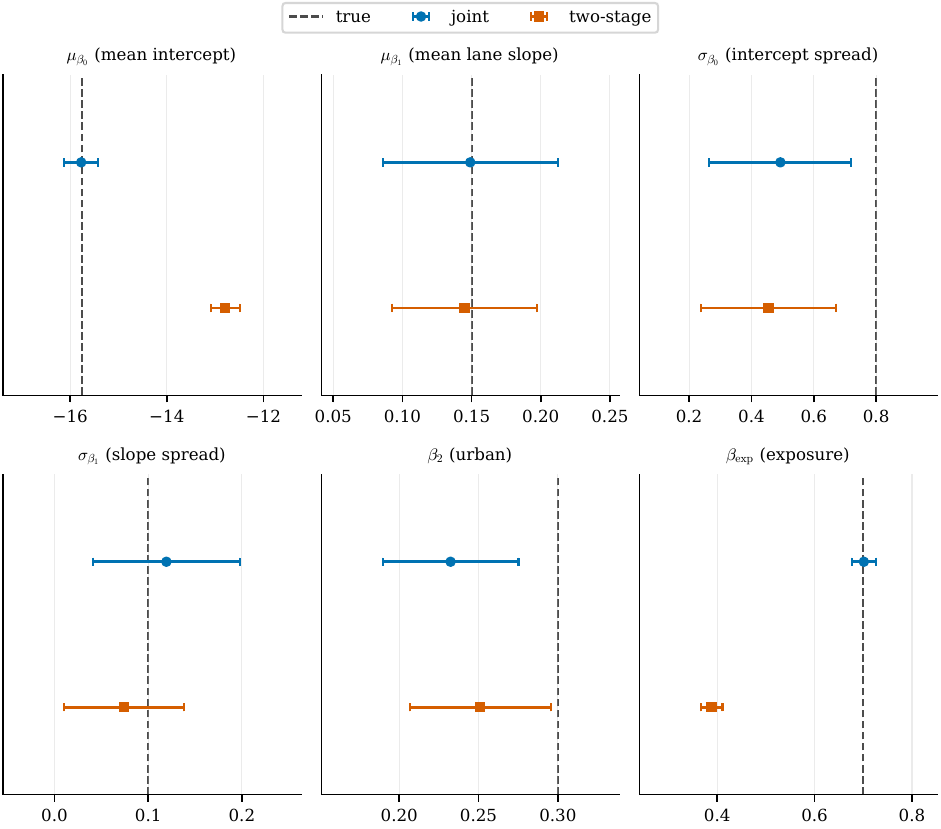}
    \caption{Parameter recovery on simulated data, one self-scaled panel per
    parameter (the intercept and slopes live on incompatible scales). Dashed
    line: true value; markers: joint and two-stage posterior means with
    $\pm 1$ SD. The joint estimates sit on the truth; the two-stage exposure
    exponent $\beta_{\mathrm{exp}}$ (lower right) is pulled far below it, the
    attenuation signature of severed crash-to-ADT feedback.}
    \label{fig:sim}
\end{figure}

The joint model recovered the crash-submodel parameters within their
posterior credible intervals (\Cref{fig:sim}): the exposure exponent
at $0.701$ against a true $0.700$, with the ADT-noise scale $\sigma_{\mathrm{adt}}$ the
lone near-miss ($0.829$ vs.\ a true $0.800$, a $0.03$ overshoot on a
parameter estimated to $\pm 0.015$). The two-stage pipeline did not: with
a true $\beta_{\mathrm{exp}} = 0.7$, its exposure exponent collapsed to
$0.389$, a bias of $-0.311$, and the intercept $\mu_{\beta_0}$ was
biased by $+2.96$, both far outside the joint model's intervals; worse, the
two-stage exponent's own credible interval is narrow and excludes the truth,
so the procedure is not merely biased but confidently wrong. The signs
are the ones the attenuation argument predicts: the frozen, covariate-only
imputations dilute the exposure signal, pulling $\beta_{\mathrm{exp}}$
\emph{below} its true value despite a prior centered at $1$ that would
otherwise pull it up, while the intercept rises to absorb the slack and
hold the mean prediction in place.

Joint inference at full scale is computationally infeasible: 408{,}304
segments include 280{,}283 latent ADT variables, more than NUTS can
sample with practical convergence. We fit on a stratified subsample of
$120{,}468$ segments containing $10{,}000$ latent ADT variables. The
stratification is by data-availability type, not by a target total. We
keep \emph{every} segment with both measured ADT and crashes
($100{,}468$ of them), which carry observed exposure and add no latent
parameters, and draw $10{,}000$ each (under a fixed seed) from the
missing-ADT-with-crashes and zero-crash strata. The cap binds only on the
missing-ADT-with-crashes group: each such segment contributes one sampled
latent ADT variable, and that count, not the number of segments, is
what governs NUTS cost, with $10{,}000$ the largest we could sample at
practical convergence. The zero-crash draw is held to the same size so the
informative-zero stratum is represented without overwhelming the
likelihood. All three segment types are retained because zeros are
informative data, not missing data: a segment with no crashes in 13 years
is evidence that it is low-risk. Excluding zeros would condition on
$\text{crashes} > 0$ and bias the partial pooling \citep{hauer2002eb}.
The ADT submodel is also fit on the full $128{,}021$ measured segments
separately to provide informed priors for the latent ADT in the joint
run.

\section{Results}
\label{sec:results}

\subsection{Convergence and Diagnostics}
\label{sec:convergence}

Both stages converged cleanly: in the joint model the maximum $\hat R$
is $1.01$ and there were no divergent transitions. The slowest-mixing
parameters are the per-FC exposure exponents
$\beta_{\mathrm{exp},\, \mathrm{fc}}$ and group intercepts
$\beta_{0,\, \mathrm{fc}}$, at bulk ESS $\approx 540$ (the expected
intercept--slope correlation within a class), while the length exponent
$\beta_{\mathrm{len}}$ and the exposure hypermean mix far faster (bulk ESS
$> 7{,}000$) and the latent ADT variables faster still. Pareto-$k$
diagnostics are good for $99.9\%$ of observations; the remaining $124$
high-leverage segments are characterized in \Cref{sec:limitations}
rather than removed.

\subsection{Posterior Estimates and Sublinear Exposure}
\label{sec:posteriors}

\begin{figure}[t]
    \centering
    \includegraphics[width=\linewidth]{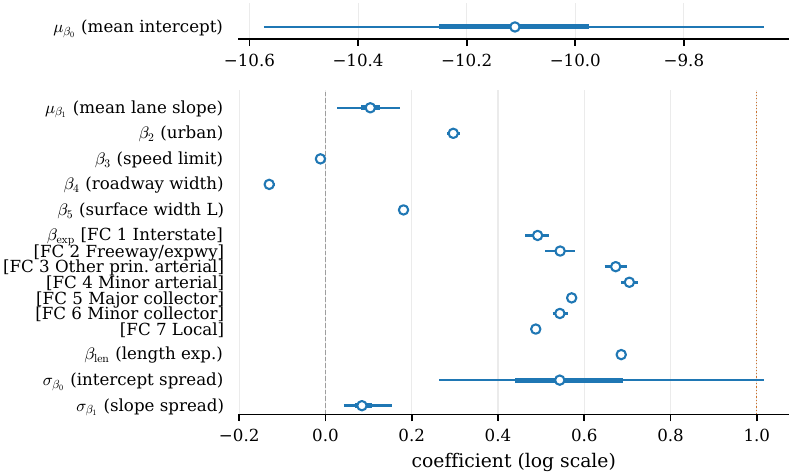}
    \caption{Posterior coefficient estimates with 94\% HDIs. The log-rate
    intercept $\mu_{\beta_0}$ (top panel) is drawn on its own scale; the
    effect-size coefficients (bottom panel) share a common axis with
    reference lines at $0$ (no effect, dashed) and $1$ (proportional
    exposure, dotted). The seven $\beta_{\mathrm{exp}}$ entries are the per-FC
    exposure exponents (FC 1--7); they and the length exponent
    $\beta_{\mathrm{len}}$ all sit well below $1$, sublinear in both
    traffic and length.}
    \label{fig:forest}
\end{figure}

The headline finding is the exposure structure. Crash count scales
\emph{sublinearly} with traffic in every functional class (the
safety-in-numbers effect documented by \citet{hauer2001overdispersion}),
but the strength of the effect varies by class: the per-FC exposure exponent
$\beta_{\mathrm{exp},\, \mathrm{fc}}$ ranges from $\betaExpMin$ (interstates
and local roads, the flattest) to $\betaExpMax$ (arterials, the steepest),
all well below $1$ (\cref{fig:forest}). A single pooled exponent would sit at
$\betaExpScalar$ and average this gradient away. Crash count also grows sublinearly with
segment \emph{length}: the estimated length exponent is $\beta_{\mathrm{len}}
= \betaLen$ (94\% HDI $[\betaLenLo, \betaLenHi]$), not the proportionality
($\beta_{\mathrm{len}} = 1$) a fixed offset would impose. We arrived at this
class-varying, length-estimated exposure structure by model checking
(\cref{sec:ppc}); we do not claim universality, as the exponents are
posteriors on Ohio 2013--2025 inventory segments.

The covariate effects sharpen the picture without overturning it
(\Cref{fig:forest}). Urban
segments carry $\beta_2 = +0.503$ on the log scale (94\% HDI
$[0.488, 0.520]$), about $65\%$ higher crash count at fixed exposure,
a known but often unquantified urban premium. Higher posted speed limits
are associated with \emph{fewer} crashes per segment,
$\beta_3 = -0.090$ (HDI $[-0.098, -0.083]$); this is the access-control
signal (high-speed roads are typically grade-separated or have fewer
driveways), not a claim about severity, where the sign reverses.
Wider total roadways have fewer crashes, $\beta_4 = -0.072$
(HDI $[-0.087, -0.059]$), consistent with a shoulder and clear-zone
effect.

The $\beta_5 = +0.200$ on \texttt{SURFACE\_WIDTH\_LEFT} (HDI
$[0.189, 0.212]$) is the result that requires explanation rather than
defense. The variable measures the paved width on the left of the
centerline; in Ohio's inventory, large values flag divided roads with
painted or paved medians. Segments with high left surface width are
therefore typically divided arterials with frequent left-turn movements
at signalized intersections, high-conflict geometry. The mechanism
is not ``wider pavement causes crashes'' but that the variable acts as
a proxy for road type the inventory does not encode directly. We treat
the sign as an empirical association, not a design recommendation.

Adding the four fixed-effect covariates to the partial-pooling model
narrowed the between-class spread parameters: $\sigma_{\beta_0}$ posterior
mean fell from $0.65$ (3-covariate model) to $0.60$ (6-covariate), and
$\sigma_{\beta_1}$ from $0.14$ to $0.12$. Variance previously attributed
to between-FC differences is now explained by within-FC covariates.

\subsection{Model Comparison}
\label{sec:loo}

We compare model variants by PSIS-LOO \citep{vehtari2017loo}
(\cref{tab:loo}): the headline model with a per-FC exposure exponent and an
estimated length exponent (\cref{sec:ppc}); its scalar-exponent,
fixed-offset predecessor; the 3-covariate partial-pooling model; and
complete-pooling counterparts at both covariate counts. Each comparison
below isolates one modeling axis.

\begin{table}
    \centering
    \caption{PSIS-LOO comparison of model variants on the joint-model
    subsample. Lower $\mathrm{elpd}_{\mathrm{loo}}$ indicates a worse model;
    $p_{\mathrm{loo}}$ is the effective parameter count. Pareto-$k$ counts
    segments with $k > 0.7$. The two complete-pooling rows use the
    scalar-exposure, fixed-offset crash structure, so the partial-vs-complete
    contrast is made on a common structure; the exposure relaxation is
    assessed separately in the top two rows.}
    \label{tab:loo}
\begin{tabular}{lrrrr}
    \toprule
    Model & $\mathrm{elpd}_{\mathrm{loo}}$ & SE & $p_{\mathrm{loo}}$ & Pareto-$k > 0.7$ \\
    \midrule
    6-cov, partial pooling               & $-391{,}783$ & $704$ & $1{,}427$ & $151$ \\
    3-cov, partial pooling               & $-392{,}768$ & $701$ & $1{,}466$ & $177$ \\
    6-cov, complete pooling              & $-396{,}563$ & $728$ & $41$ & $1$ \\
    3-cov, complete pooling              & $-399{,}600$ & $709$ & $12$ & $0$ \\
    \bottomrule
\end{tabular}

\end{table}

\emph{Pooling.} On a common crash structure, partial pooling beats complete
pooling by $\Delta\mathrm{elpd} = \loocpDelta$ (SE $\loocpDeltaSE$), a margin
much larger than its uncertainty. \emph{Covariates.} The 6-covariate model
beats the 3-covariate version by $\loocovDelta$ elpd with a \emph{lower}
effective parameter count ($\looPlooSix$ vs.\ $\looPlooThree$): the added
covariates carry independent signal, not just complexity. \emph{Exposure
structure.} Relaxing the fixed exposure assumptions, replacing one
scalar and a fixed offset with a per-FC exponent and an estimated length
exponent, improves elpd by a further $\loovTwoDelta$ (SE $\loovTwoDeltaSE$),
the largest single gain in the table, while \emph{reducing} both the
effective parameter count ($\looPlooVTwo$ vs.\ $\looPlooSix$) and the
high-$k$ count ($\looKVTwo$ vs.\ $\looKSix$). It is a strictly better fit,
not a more flexible one; \cref{sec:ppc} shows the posterior predictive checks
that motivated it.

We do not include a no-pooling baseline. With $7$ well-populated
functional class groups, independent per-FC fits would be a strawman;
the substantive question is whether the data support sharing strength
across classes (partial vs.\ complete pooling), and the answer is
unambiguous.

\subsection{Posterior Crash Rates}
\label{sec:rates}

\begin{figure}[t]
    \centering
    \includegraphics[width=\linewidth]{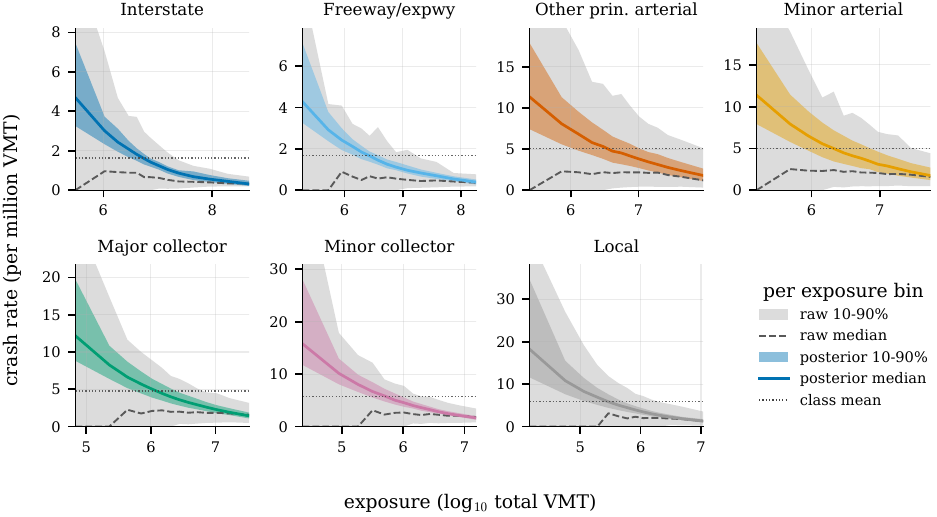}
    \caption{Shrinkage as a funnel: crash rate (per million VMT) against
    exposure (log$_{10}$ total VMT), by functional class. Within each
    exposure bin the grey band spans the 10th to 90th percentile of the raw
    rate (observed crashes divided by VMT) and the dashed line its median;
    the coloured band and solid line are the same summaries for the model's
    posterior rate, and the dotted line marks the class mean. At low
    exposure the raw rate is a broad funnel of mostly noise: its 10th to
    90th percentile spread collapses by a factor of 4 to 23 across exposure,
    converging onto the narrow, stable posterior band where data are rich.
    That collapse is regression to the mean made visible, the posterior
    trusting a segment's own rate only in proportion to its exposure. It is
    the Bayesian analogue of Hauer's EB weighting, \cref{eq:eb}, with the
    weight learned rather than fixed.}
    \label{fig:shrinkage}
\end{figure}

Posterior mean crash rates per million vehicle miles travelled (VMT),
defined as predicted crashes divided by actual VMT and averaged within each
functional class, are shown in \cref{tab:rates}, with \cref{fig:shrinkage} showing
the segment-level shrinkage behind them. The rates split by access control
rather than ordering monotonically down the functional hierarchy:
limited-access facilities are far safer per mile travelled (interstates at
$\rateInterstate$, freeways/expressways comparable) than every other class.
The remaining classes run several times higher and cluster together with no
clean ranking among themselves (for example, other principal arterials at
$\rateOtherArterial$ and local roads, the highest, at $\rateLocal$;
\cref{tab:rates}). The
within-class spread is wide (a handful of segments per class run several
times the class mean), which is exactly the segment-level heterogeneity
the partial-pooling crash submodel and the per-FC exposure exponent
estimate rather than assume away.

\begin{table}
    \centering
    \caption{Posterior mean crash rate per million VMT by functional class,
    on measured-ADT segments, with within-class standard deviation, minimum,
    and maximum across segments. The rate is predicted crashes divided by
    actual vehicle-miles travelled, exposure-normalized and invariant to
    the exposure parameterization. Rates are point posterior means of
    segment-level distributions; per-segment uncertainty is omitted for
    compactness.}
    \label{tab:rates}
\begin{tabular}{rlrrrr}
    \toprule
    FC & Class & Mean & SD & Min & Max \\
    \midrule
    1 & Interstate                 & 1.63 & 1.43 & 0.13 & 27.3 \\
    2 & Freeway / expressway       & 1.67 & 1.30 & 0.19 & 21.5 \\
    3 & Other principal arterial   & 5.10 & 3.29 & 0.54 & 66.6 \\
    4 & Minor arterial             & 4.95 & 3.19 & 0.50 & 43.8 \\
    5 & Major collector            & 4.77 & 3.69 & 0.42 & 151.0 \\
    6 & Minor collector            & 5.74 & 4.97 & 0.85 & 92.5 \\
    7 & Local                      & 5.94 & 6.22 & 0.36 & 233.5 \\
    \bottomrule
\end{tabular}

\end{table}

\subsection{Posterior Predictive Checks and Model Expansion}
\label{sec:ppc}

The class-varying, length-estimated exposure structure of \cref{eq:crash}
was not assumed; it was arrived at by checking a simpler model and
expanding where the checks failed, in the manner of \citet{gelman2020workflow}.
We describe that path here, because it both motivates the structure and
demonstrates the paper's thesis in miniature.

\begin{figure}[t]
    \centering
    \includegraphics[width=\linewidth]{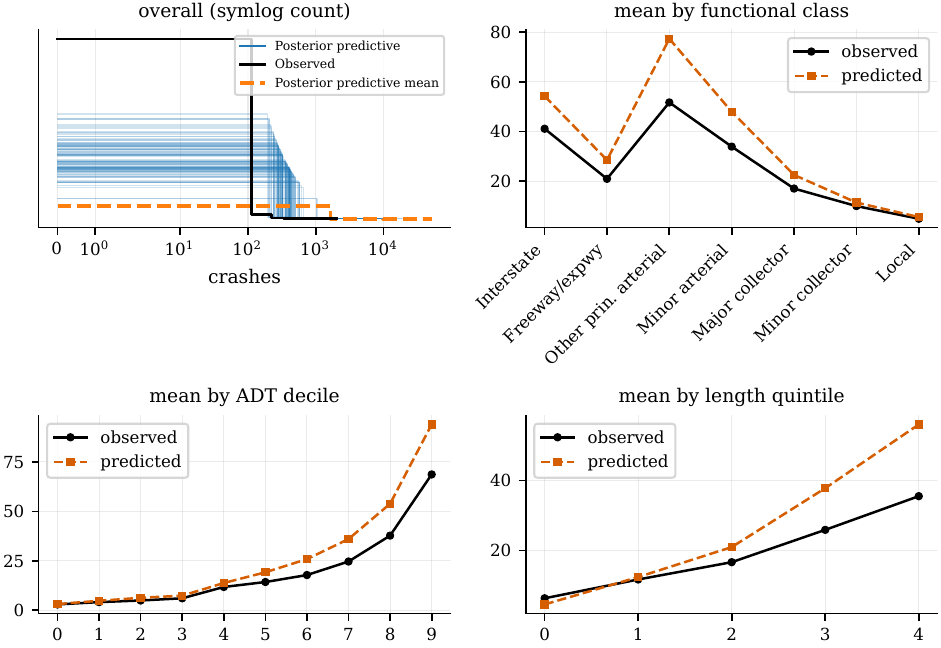}
    \caption{Standard-model posterior predictive checks (single exposure
    exponent, fixed length offset). (i) Overall count distribution on a
    symmetric-log scale; (ii)--(iv) observed versus posterior-predictive mean
    counts by functional class, ADT decile, and segment-length quintile. The
    per-class means track, but predicted means over-predict by $1.4$--$1.6
    \times$ in the upper ADT deciles and the longest-segment quintile,
    the misfit that motivates the exposure expansion.}
    \label{fig:ppc}
\end{figure}

\paragraph{The standard model and its misfit.} We began with the field-standard
specification: a single exposure exponent $\beta_{\mathrm{exp}}$ and segment
length as a fixed offset (exponent $1$), exactly the form an HSM-style safety
performance function takes. Posterior predictive checks compare simulated
crash counts to observed counts along four marginal views: the overall count
distribution (log scale, for the heavy zero mass and long tail), and means by
functional class, by ADT decile, and by segment-length quintile. The overall
distribution and the per-class means track the data, but the two
exposure-facing panels reveal a systematic over-prediction that grows at the
tails: predicted means run about $1.4$--$1.6\times$ observed in the upper ADT
deciles and the longest-segment quintile (\Cref{fig:ppc}).

\paragraph{Diagnosis.} The two tail misfits are not redundant. Length and ADT
are negatively correlated across segments (long segments are
disproportionately rural and low-traffic), so a single confounded effect
would bias the two panels in opposite directions, not the same one. Instead
each panel implicates a distinct fixed assumption: the ADT-decile bias points
at the single scalar exponent, which cannot bend across the traffic range,
and the length-quintile bias points at the exponent-one offset, which
over-credits exposure on long, homogeneous, low-conflict rural mainline.

\begin{figure}[t]
    \centering
    \includegraphics[width=\linewidth]{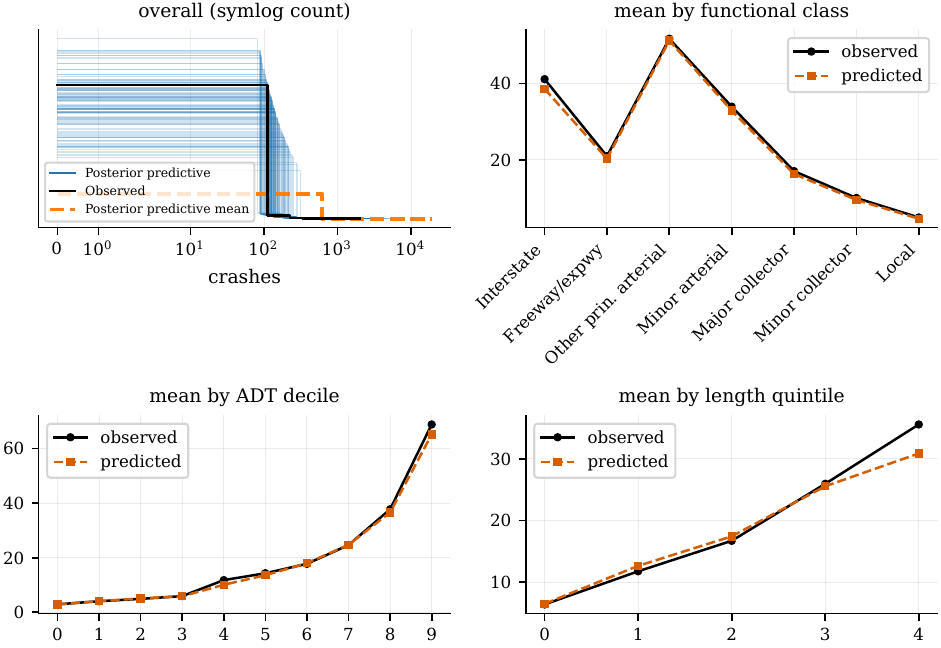}
    \caption{Expanded-model posterior predictive checks (per-FC exposure
    exponent, estimated length exponent), same four views as
    \cref{fig:ppc}. Predicted and observed means now coincide across
    functional class, ADT decile, and segment-length quintile; the tail
    over-prediction is resolved.}
    \label{fig:ppcvtwo}
\end{figure}

\paragraph{Expansion and re-check.} We relaxed exactly those two assumptions,
letting $\beta_{\mathrm{exp}}$ vary by functional class under a shared
hyperprior and estimating the length exponent $\beta_{\mathrm{len}}$
(\cref{eq:crash}), the same partial-pooling move the model already applies
to the intercept and lane slope. The expansion is decisively supported:
$\Delta\mathrm{elpd} = \loovTwoDelta$ (SE $\loovTwoDeltaSE$) over the standard
model (\cref{tab:loo}), with a \emph{lower} effective parameter count and
fewer high-leverage segments: a better fit, not merely a more flexible one.
The data revise both exponents below proportionality
($\beta_{\mathrm{len}} = \betaLen$; per-FC $\beta_{\mathrm{exp}}$ from
$\betaExpMin$ to $\betaExpMax$, \cref{fig:forest}), and the two are cleanly
identified: their posterior correlation is below $0.1$ despite the
length--ADT collinearity, so this is a genuine decomposition rather than a
trade-off. Re-running the checks on the expanded model closes the gap:
predicted and observed means coincide across all three groupings, and
within-class calibration tightens to within a few percent of parity
(\Cref{fig:ppcvtwo}).

\subsection{Held-Out Year Validation}
\label{sec:heldout}

The 13-year crash window assumes road characteristics and crash-reporting
practice are approximately stationary. A natural check, which we leave to
future work, refits the joint model on $2013$--$2022$ and evaluates
predictive performance on $2023$--$2025$, asking whether
$\beta_{\mathrm{exp}}$ and the FC-level mean rates are stable when held-out
years are dropped from fitting. Stability would not prove stationarity but
would weaken the argument against it; we expect it to be a more substantive
robustness check than a no-pooling baseline.

\section{Discussion}
\label{sec:discussion}

\subsection{What the Model Relaxes}
\label{sec:generalization}

\Cref{tab:correspondence} maps each fixed-parameter assumption from the
EB recipe to its relaxation in the present model. Read down the
left column for the EB-fluent description and across to the right for
the change.

\begin{table}
    \centering
    \caption{Correspondence between Empirical Bayes fixed-parameter
    assumptions and the present model's relaxations.}
    \label{tab:correspondence}
    \begin{tabular}{p{0.45\textwidth} p{0.5\textwidth}}
        \toprule
        EB assumption & Bayesian relaxation \\
        \midrule
        SPF coefficients $(\alpha, \beta)$ are MLE point estimates fit at
        calibration & Joint posterior over all coefficients, propagated
        end-to-end into segment rates \\
        Overdispersion $\phi$ is one number per road type & Per-FC
        overdispersion $\alpha_{\mathrm{fc}}$ estimated jointly with the
        SPF coefficients \\
        ADT is observed & Latent variable where missing, joint with the
        crash likelihood so counts inform exposure \\
        Exposure exponent fixed at SPF calibration & Estimated and partially
        pooled across functional class ($\beta_{\mathrm{exp},\,\mathrm{fc}}$
        from $\betaExpMin$ to $\betaExpMax$); segment length likewise carries
        an estimated exponent $\beta_{\mathrm{len}} = \betaLen$ \\
        Output is the point estimate $\hat{\lambda}^{\mathrm{EB}}$ &
        Full posterior $p(\lambda_i \mid \text{data})$ per segment \\
        \bottomrule
    \end{tabular}
\end{table}

\subsection{When Empirical Bayes Remains Appropriate}
\label{sec:eb-still-good}

EB remains the right tool when its assumptions hold. Where ADT is broadly
observed, where road types within a class are homogeneous, where point
estimates suffice for the application, and where practitioner
accessibility outweighs uncertainty quantification, EB delivers most of
the same shrinkage at a fraction of the computational and methodological
cost. The present model adds value where these conditions strain, in
particular when ADT is missing on a majority of segments and the
downstream application consumes a distribution rather than a number. We
mean this seriously: the spreadsheet workflow EB enables is a feature of
the method that no MCMC sampler will replace for many of the audiences
EB serves.

\subsection{Limitations}
\label{sec:limitations}

Several limitations bear stating prominently rather than burying.

The ADT submodel shows regression-to-the-mean bias at the tails of the
ADT distribution: low-ADT segments are over-predicted and high-ADT
segments under-predicted, as expected at $R^2_{\log} = 0.76$. The bias
is smaller than that of a LightGBM on the same continuous features (residual
range $[-1.14, +0.37]$ vs.\ $[-1.31, +0.71]$ on the log scale) but real. Wider
posteriors on extreme segments propagate the uncertainty honestly; they
do not eliminate the bias. Closing it would require features not
available in the inventory, in particular access control and intersection
density.

The exposure expansion of \cref{sec:ppc} resolves the tail over-prediction
in aggregate, but a second-order residual remains: split by urban/rural, the
expanded model still over-predicts mean crashes by roughly $20$--$30\%$ on
rural \emph{mid}-length segments, even as it calibrates well on urban
segments and at both length extremes. The likely cause is the same missing
covariates noted above (access control and intersection density), which
are concentrated on exactly those roads. A length exponent that varied by
class, or an explicit length-by-class interaction, could absorb it; we leave
that to future work rather than add structure the diagnostics only weakly
demand.

A small set of segments, $124$ of $\sim 120{,}000$ ($0.10\%$), have
Pareto-$k > 0.7$, indicating the LOO importance weights are heavy-tailed
and the elpd contribution is unreliable for those points. We do not
remove them; they are the cases where the model is least sure, and the
correct response is to characterize them rather than paper over them.
The high-$k$ group is overwhelmingly local roads (FC 7) at higher ADT
and crash counts than the subsample, concentrated at short-to-mid
lengths and lacking the longest segments, the regime where the inventory
and crash records together provide the thinnest signal.

The covariate set excludes three variables that the Highway Safety Manual
emphasizes (access control, median type, and intersection density)
because they are not in the inventory extract. The model estimates
predictive associations within functional class and the available
covariates; it should not be read as an estimate of pure causal effects.
The new covariates ($\beta_3, \beta_4, \beta_5$) partially substitute
for these missing differentiators but do not replace them.

ADT is assumed missing at random conditional on the road attributes in
the ADT submodel. This is plausible (DOT counting programs prioritize
higher-class roads, and functional class is in the model) but
unverifiable. Residual selection within functional class would bias the
latent ADT estimates, and consequently $\beta_{\mathrm{exp}}$, in the
direction of the selection.

The sign on $\beta_5$ (left surface width) requires domain validation.
We have offered a mechanism (proxy for divided-road geometry); a Highway
Safety Manual specialist may have a cleaner interpretation. We do not
adjust the model on the strength of an interpretation alone.

The data come from a single state and a 13-year window. Replication on
another DOT's inventory is the cleanest test of generalizability and is
left as future work.

\subsection{Integration with Risk-Aware Routing}
\label{sec:routing}

The output of this model is a posterior crash rate distribution per
inventory segment. In the routing framework of \citet{skaug2026cav},
percentile-based risk scores are replaced by posterior means and
route-level aggregation can report credible intervals rather than point
estimates, so route-level risk in this pipeline carries uncertainty bands
rather than single numbers. The integration with
the GraphHopper-based routing pipeline is in progress; we defer the
corridor demonstration to a routing-focused follow-up to keep the present
paper's scope on the Bayesian model itself. Related in-vehicle
safety systems developed alongside this routing work, covering road-risk
awareness and sun-glare avoidance for semi- and fully autonomous driving,
are described in pending patents \citep{nojoumian2025ras, nojoumian2025sas}.

\section{Conclusion}
\label{sec:conclusion}

We presented a Bayesian hierarchical model that moves beyond Hauer's
Empirical Bayes procedure by relaxing each of its fixed-parameter
assumptions in a single joint inference. Model checking drove the central
refinement: posterior predictive checks of a fixed-exposure model exposed a
tail misfit, and relaxing the exposure structure to a per-functional-class
exposure exponent and an estimated length exponent resolved it. Crash
count is sublinear in both traffic and length, the safety-in-numbers regime,
but now as class-varying posterior intervals rather than chosen values. Each segment receives a posterior crash rate distribution
that propagates ADT uncertainty end-to-end, suitable for downstream
consumers, including routing applications, that benefit from
intervals over points. EB remains appropriate where its assumptions hold;
the contribution here is computational evolution, not correction.

\section*{Data Availability}

The road inventory, traffic count segments, and crash records used in this
study are public datasets from the Ohio Department of Transportation. The
model is specified in full in \Cref{sec:model}.

\section*{Acknowledgments}

The author thanks Mehrdad Nojoumian, who advised the prior work
(\citealp{skaug2026cav}) on which this paper builds.


\end{document}